\PassOptionsToPackage{unicode}{hyperref}
\PassOptionsToPackage{hyphens}{url}
\PassOptionsToPackage{full}{textcomp} 
\documentclass[
  10pt,
  a4paper,
]{article}
\usepackage{xcolor}
\usepackage{amsmath} 
\usepackage[T1]{fontenc}
\usepackage[utf8]{inputenc}
\usepackage{textcomp}
\usepackage{lmodern}
\IfFileExists{upquote.sty}{\usepackage{upquote}}{}
\IfFileExists{microtype.sty}{
  \usepackage[]{microtype}
  \UseMicrotypeSet[protrusion]{basicmath} 
}{}
\makeatletter
\@ifundefined{KOMAClassName}{
  \IfFileExists{parskip.sty}{%
    \usepackage{parskip}
  }{
    \setlength{\parindent}{0pt}
    \setlength{\parskip}{6pt plus 2pt minus 1pt}}
}{
  \KOMAoptions{parskip=half}}
\makeatother
\usepackage{longtable,booktabs,array}
\usepackage{calc} 
\usepackage{etoolbox}
\makeatletter
\patchcmd\longtable{\par}{\if@noskipsec\mbox{}\fi\par}{}{}
\makeatother
\IfFileExists{footnotehyper.sty}{\usepackage{footnotehyper}}{\usepackage{footnote}}
\makesavenoteenv{longtable}
\usepackage{graphicx}
\makeatletter
\newsavebox\pandoc@box
\newcommand*\pandocbounded[1]{
  \sbox\pandoc@box{#1}%
  \Gscale@div\@tempa{\textheight}{\dimexpr\ht\pandoc@box+\dp\pandoc@box\relax}%
  \Gscale@div\@tempb{\linewidth}{\wd\pandoc@box}%
  \ifdim\@tempb\p@<\@tempa\p@\let\@tempa\@tempb\fi
  \ifdim\@tempa\p@<\p@\scalebox{\@tempa}{\usebox\pandoc@box}%
  \else\usebox{\pandoc@box}%
  \fi%
}
\def\fps@figure{htbp}
\makeatother
\providecommand{\tightlist}{%
  \setlength{\itemsep}{0pt}\setlength{\parskip}{0pt}}
\usepackage{stix2}
\usepackage[scaled=0.80]{beramono}
\usepackage{newunicodechar}
\newunicodechar{·}{\textperiodcentered{}}  
\newunicodechar{—}{\textemdash{}}
\newunicodechar{–}{\textendash{}}
\newunicodechar{‘}{\textquoteleft{}}
\newunicodechar{’}{\textquoteright{}}
\newunicodechar{“}{\textquotedblleft{}}
\newunicodechar{”}{\textquotedblright{}}
\newunicodechar{…}{\textellipsis{}}
\newunicodechar{→}{\ensuremath{\rightarrow}}
\newunicodechar{←}{\ensuremath{\leftarrow}}
\newunicodechar{×}{\ensuremath{\times}}
\newunicodechar{≤}{\ensuremath{\leq}}
\newunicodechar{≥}{\ensuremath{\geq}}
\newunicodechar{±}{\ensuremath{\pm}}
\usepackage[margin=1.12in]{geometry}
\usepackage{microtype}
\usepackage{booktabs,longtable,array}
\usepackage{etoolbox}
\AtBeginEnvironment{longtable}{\footnotesize\setlength{\tabcolsep}{4pt}}
\usepackage[font=small,labelfont=bf,margin=0.6cm]{caption}
\usepackage{titlesec}
\titleformat*{\section}{\large\bfseries}
\titleformat*{\subsection}{\normalsize\bfseries}
\titlespacing*{\section}{0pt}{1.4em}{0.6em}
\titlespacing*{\subsection}{0pt}{1.1em}{0.4em}
\usepackage{xcolor}
\definecolor{accent}{HTML}{1f3864}
\usepackage{hyperref}
\usepackage{xurl}
\hypersetup{colorlinks=true,linkcolor=accent,urlcolor=accent,citecolor=accent}
\usepackage{graphicx}
\setkeys{Gin}{width=\linewidth,keepaspectratio}
\usepackage{fancyhdr}
\fancypagestyle{plain}{\fancyhf{}\fancyfoot[C]{\footnotesize\thepage}}
\renewenvironment{abstract}
 {\begin{center}\begin{minipage}{0.88\textwidth}\small
  \setlength{\parindent}{0pt}\setlength{\parskip}{0.45em}
  \begin{center}\textbf{Abstract}\end{center}\vspace{0.1em}}
 {\end{minipage}\end{center}\vspace{1.0em}}
\usepackage{framed}
\renewenvironment{quote}
 {%
  \MakeFramed{\advance\hsize-\width\FrameRestore}\small\vspace{0.15em}}
 {\vspace{0.15em}\endMakeFramed}
\usepackage{bookmark}
\IfFileExists{xurl.sty}{\usepackage{xurl}}{} 
\hypersetup{
  hidelinks,
  pdfcreator={LaTeX via pandoc}}

\author{}
\date{}

\begin{document}

\thispagestyle{plain}
\begin{center}
{\LARGE\bfseries From Traceability to Justifiability:\\[0.25em] Accountability Structures in Agentic Software Engineering\par}
\vspace{1.3em}
{\large Rashid Azarang\par}
\vspace{0.25em}
{\small\itshape Independent Researcher, San Pedro Garza Garc\'ia, Nuevo Le\'on, Mexico\par}
{\small \href{mailto:rashid@mentu.ai}{rashid@mentu.ai} \,\textperiodcentered\, \href{https://rashidazarang.com}{rashidazarang.com} \,\textperiodcentered\, ORCID \href{https://orcid.org/0009-0008-5528-4246}{0009-0008-5528-4246}\par}
\vspace{0.4em}
{\small Preprint \,\textperiodcentered\, 2026-08-21\par}
\end{center}
\vspace{0.4em}

\begin{abstract}

When a pipeline promotes an AI system, its published records make a
claim: that the thing evaluated is the thing deployed, and that the
evidence licensed the transition. This paper measures, at scale and from
public material only, whether the records can even express that claim
and whether the claim is realized where it is declared. Two instruments
carry the measurement. The first is a two-class documentation survey of
47 delivery platforms (20 CI/CD, 27 model-serving and agent platforms)
under one fixed three-label protocol, every grading performed twice with
the second pass blind, every consulted page pinned by content hash and
fetch date. Its central finding, stated as the bounded claim it is:
across the 47 platforms, in a survey of 188 double-graded cells, we
found no platform whose default record emits a content-addressed
identity of the behavioral tuple (model version, instructions, tool
definitions, retrieval and runtime configuration),
a bound the blind pass supports unaided, grading that column default on
zero of the 47 platforms, while immutable nominal versioning of the
tuple is arriving as the
agent platforms' default answer, on 16 of 27 platforms: version
integers behind mutable pointers, the layer the artifact supply chain
community already found insufficient. The second instrument computes a
pipeline's realized assurance depth from its published exhaust alone
and compares it to the depth its configuration declares, without asking
anyone. Applied to a frozen two-stratum frame of 30 public repositories
graded twice from a hashed raw archive (the second pass blind;
cell-level agreement 23 and 19 of 30, with both passes independently
finding the same five full realizations before any reconciliation), it
measures how much of the declared assurance the public surface
realizes, and its sharpest result is a verifiability hole: seven of the
15 repositories selected precisely because they adopt attestation
tooling publish source-only releases, so the binding their workflows
declare cannot be checked on the surface where they declare it: the
declaration and the verifiable evidence live on different public
surfaces with nothing linking them. Where the declared surface can be
checked, it mostly checks out: five of the seven measurable adopters
realize the declared binding end to end (one third of the stratum), and
the two measured shortfalls both fall at the identity-binding rung.
Together the two results locate the field's records structurally short
of justifiability, the one level of the traceability ladder that can
refuse a transition, and they time-stamp the claim: the survey carries
an expiry clock, and the paper states what would falsify each finding.


\smallskip
\noindent {\footnotesize Every quantity in this paper is descriptive. The program's confirmatory arm is a prospective second-site study, registered at OSF (DOI \href{https://doi.org/10.17605/OSF.IO/D3JFV}{10.17605/OSF.IO/D3JFV}) and deliberately not reported here (Section 9).}

\end{abstract}

\subsection{1. Introduction}\label{introduction}

A modern agentic delivery pipeline, built with a blocking behavioral
gate and an evaluation suite, authorized a promotion on evidence that
could not say what it described: the identity field was null in all
fourteen evidence records of the build, and the one identifier that
survived named a runtime slot verified identical across two different
releases. Nothing in the pipeline's own checks flagged the absence; it
was detected by an outside instrument reading the pipeline's published
artifacts, and the defect, a contingent wiring error, was repaired
within the week. What the case exposes is not one estate's carelessness
but a pair of measurable questions this paper takes to the field: what
can a delivery platform's \emph{default} records express about a
promotion, and how much of the assurance a pipeline \emph{declares} does
its published evidence \emph{realize}?

The joint contribution, stated once and stated as the two rungs it
actually spans: \textbf{the survey and the instrument measure the same
ladder at its two ends. The survey establishes the ceiling: no pipeline
can declare, let alone realize, behavioral-rung assurance today, because
no platform's default record expresses the identity that rung needs. The
depth study measures the floor: how much of even artifact-layer declared
assurance the published evidence realizes, measured precisely where the
needed record class does exist and is default.} The halves are not
explanation and instance; the survey does not cause the measured gap,
and the paper claims no such link. What each supplies the other is
scope: the survey without the instrument is documentation criticism with
no demonstration that realization is measurable at all; the instrument
without the survey would invite the false reading that the behavioral
rung is one more adoption push away, when the record class it needs does
not exist to adopt. Between ceiling and floor sits a third result
neither half predicted: even where the records exist, the declaration
and the verifiable evidence can live on surfaces nothing links (Section
7.1).

\subsubsection{1.1 Three levels, and the standards' own
promise}\label{three-levels-and-the-standards-own-promise}

Traceability, as the standards define it, is not a modest property.
ISO/IEC/IEEE 12207 defines it as the ``degree to which a relationship
can be established among two or more logical entities, especially
entities having a predecessor-successor relationship to one another,
such as requirements, system elements, verifications, or tasks''
{[}\hyperref[ref-iso12207]{27}{]}. A configuration-management technical
report, indexed as a further sense of the same headword in the field's
consolidated vocabulary, defines it as the ``degree to which each
element in a software development product establishes its reason for
existing'' {[}\hyperref[ref-isotr18018]{28}{]}. NIST fuses the concepts
this paper's title separates into one design principle: control
enhancement SA-8(22), \emph{Accountability and Traceability}, ``states
that it is possible to trace security-relevant actions (i.e.,
subject-object interactions) to the entity on whose behalf the action is
being taken'' {[}\hyperref[ref-nistsp80053r5]{34}{]}. On their own terms
these definitions already reach toward justification, and this paper
takes them at their word: it moves \emph{from} traceability, not against
it. What no instrument in the traced record does is \emph{adjudicate}
what the definitions promise. Read in software supply chain terms, which
is the community whose evidence classes this paper measures, the same
movement runs one layer up the chain: the supply chain community solved
identity for artifacts with content addressing, is watching the same
problem recur for models, and this paper measures the link where the
chain now ends, the behavioral configuration a delivery platform records
when an AI system ships.

Three levels organize the measurement. \textbf{Provenance} asks where a
thing came from; W3C PROV {[}\hyperref[ref-provdm2013]{39}{]} and
OpenLineage {[}\hyperref[ref-openlineage2026]{36}{]} own the level and
this paper contributes nothing to it. \textbf{Traceability} asks how the
things relate; organizations approximate it with ticket keys, build
numbers, and consistent identifiers. \textbf{Justifiability} asks
whether the evidence available at the time licensed the transition, and
it is the only one of the three that can return \emph{no}. The word has
neighbors, named at first use: Goal Structuring Notation carries a
Justification node type that warrants a step in an assurance argument
{[}\hyperref[ref-hawkins2011]{21},
\hyperref[ref-goodenough2015ea]{18}{]}, and recent work argues for
agentic AI whose decisions are grounded in inspectable ontological
context {[}\hyperref[ref-justified2025]{29}{]}. Neither is the predicate
defined here: a promotion-time check over already-published records that
refuses the transition when a required relationship is absent. The
distinction is quotable from the primary texts of the lower levels:
in-toto states that judging whether a supply chain layout is sound is
not its role {[}\hyperref[ref-intotospec2024]{24}{]}; PROV imposes ``no
prescriptive requirements on the nature of plans''
{[}\hyperref[ref-constraints2013]{15}{]}; GUAC ingests software metadata
and maps relationships between software {[}\hyperref[ref-guac]{20}{]},
returning answers, never refusals.

The setting is agentic software engineering
{[}\hyperref[ref-agentic2026]{5},
\hyperref[ref-semiexecutable2026]{45}{]}: AI agents author most of the
code, humans hold promotion authority, and the platform's ordinary
exhaust is the only record there is. The premise that makes the setting
interesting is cited background {[}\hyperref[ref-sculley2015]{44},
\hyperref[ref-from2026]{17}{]}: behavior depends on a tuple (source,
model version, instructions, retrieval configuration, tool definitions,
runtime configuration, environment) whose components vary independently
of the source. The premise carries its own demarcation, which bounds
every claim below: where the components cannot vary independently
(models pinned, prompts in-tree, no runtime retrieval, static tools),
the tuple collapses onto the source digest, conventional artifact
provenance is sufficient, and this paper predicts no advantage from
anything more elaborate.

\subsubsection{1.2 Contributions}\label{contributions}

\begin{enumerate}
\def\labelenumi{\arabic{enumi}.}
\item
  \textbf{A specification.} Four continuity conditions under which a
  promotion is justifiable (Section 3), assembled from cited prior art
  with two declared residues (the invariant/variant partition, Section
  3.1; authority's temporal-order limb), the nominal versus
  content-addressed distinction that cross-cuts them, and the printed
  0-4 depth rubric that operationalizes them.
\item
  \textbf{The depth instrument, public form.} A pipeline's realized
  assurance depth is computable from its published exhaust alone and
  comparable to its declared depth, with no producer cooperation, no
  interviews, and none of the signed attestations the supply chain
  frameworks presuppose (Section 4). Most CI/CD assurance measurement
  reads what is configured; this instrument measures whether the
  published evidence can support the claim the configuration implies,
  and it scales with the estate rather than with an auditor's calendar.
\item
  \textbf{The two-class default-emission survey.} 47 platforms, one
  fixed three-label protocol, every cell graded twice with the second
  pass blind, every consulted page pinned (Section 5); the sealed tables
  and the ladder of what the default record can express (Section 6).
\item
  \textbf{The measured gap, and the documentary finding that explains
  it.} On a frozen 30-repository public frame, the
  declared-versus-realized sharpest depth-study finding is a
  verifiability hole (most attestation adopters declare a binding on a
  release surface that carries no artifacts to check), most adopters
  whose surface can be checked realize their declaration (five of seven;
  one third of the stratum), and both measured shortfalls fall at the
  binding rung (Section 7.1); the survey's documentary finding, with its
  definitional corollary conceded as such and its expiry clock running,
  explains why the gap is structural: no default record carries the
  identity the declared depth needs (Section 7.2).
\end{enumerate}

\subsubsection{1.3 What would kill this
paper}\label{what-would-kill-this-paper}

Per genre. The survey's negative cells die by exhibition: every
bounded-absent grade names the consulted pages and is falsified by
naming a page that describes the capability; the survey's headline dies
by exhibiting one platform that emits a content-addressed
behavioral-tuple identity by default, and expires when the
standardization path ships one (the clock is in Section 7.2). The depth
instrument dies on its rubric: a rung determination shown false against
a repository's actual records kills the operationalization, and the
study's raw archive is published precisely so any reader can re-derive
any cell. The prevalence conjecture is declared untested at the
behavioral rung. Everything an existence proof cannot touch is carried
by the registered study (Section 9), frozen before its observation
window opened, and not answered here.

\subsection{2. What existing frameworks cannot refuse, and
why}\label{what-existing-frameworks-cannot-refuse-and-why}

Related work here has one job: to establish that no existing framework
performs the specific measurement this paper performs. The positioning
rests on a systematic adversarial literature sweep whose protocol,
search set, and verified returns are archived in the program repository.
The conceptual boundary between what artifacts decide and what requires
judgment is independently occupied in at least six literatures: Rushby
partitions assurance doubt into a logical half reducible to automated
checking and an epistemic half that ``requires human experience and
insight'' {[}\hyperref[ref-rushby2013]{41},
\hyperref[ref-rushby2015nasa]{42}{]}; Ackerman names the enduring
distance the social-technical gap {[}\hyperref[ref-ackerman2000]{2}{]};
Cook places CI/CD pipelines and repositories below the \emph{line of
representation}, where nothing ``can be seen or acted upon directly''
{[}\hyperref[ref-cook2020_atl]{16},
\hyperref[ref-woods_allspaw2020_revealing]{55}{]}; Naur made the point
for programming four decades ago {[}\hyperref[ref-naur1985]{33}{]};
Hollnagel names the divergence between record and practice
work-as-imagined versus work-as-done
{[}\hyperref[ref-hollnagel2015_whitepaper]{22}{]}. Each draws the
boundary; none measures its mechanical side from platform exhaust.

The supply chain frameworks are the strongest occupants of the
mechanical side, and the seam is precise. in-toto provides cryptographic
custody for artifacts {[}\hyperref[ref-intoto2019]{23}{]}; SLSA's threat
model treats machine-verifiable build evidence as sufficient against its
enumerated threats {[}\hyperref[ref-slsa_threats]{51}{]}; Proof-or-Stop,
in this paper's exact setting, defines advancement as an admissibility
decision that can return NO {[}\hyperref[ref-prooforstop2026]{38}{]}.
All of them verify what cooperating producers emitted under obligations
the framework imposes (sign the step, isolate the builder
{[}\hyperref[ref-slsa2023]{49}{]}); none of them asks the two questions
measured here, which assume no cooperation at all: whether the
\emph{default} record, absent any framework adoption, can express the
identity the assurance claim needs, and whether the published exhaust of
real pipelines realizes the depth their configurations declare.
Computability over published artifacts is established practice at the
provenance level {[}\hyperref[ref-macaron2023]{30},
\hyperref[ref-on2026]{35}, \hyperref[ref-sigstore2022]{47}{]}, a level
this paper concedes; the adherence-gap literature has shown
declared-versus-emitted distance to be systematic for SBOM tooling
{[}\hyperref[ref-adherencegap2026]{4},
\hyperref[ref-operationalizing2026]{37}{]} and for research software
{[}\hyperref[ref-operationalizing2026]{37}{]}; what is added here is the
promotion-time framing, the two-source construction (declared depth from
configuration, realized depth from exhaust, never from the same
records), and the survey that explains the gap's location.

One inversion from an older profession frames the method. Auditing
standards rank inquiry as the weak form of evidence, to be combined with
inspection or reperformance {[}\hyperref[ref-as1105]{10},
\hyperref[ref-isa500]{26}{]}, and treat undocumented work as undone
{[}\hyperref[ref-isa230]{25}, \hyperref[ref-as1215]{11}{]}. Both of this
paper's instruments are built to that discipline: the survey grades only
what documentation states, bounded by named pages; the depth study
grades only what the archived exhaust shows, and everything the archive
cannot support is counted as undetermined rather than resolved by asking
anyone.

Adjacent framings, named: \emph{Where Accountability Lives} maps human
responsibility to workflow artifacts {[}\hyperref[ref-where2026]{54}{]}
and its companion roadmap analyzes agent accountability in terms of
service {[}\hyperref[ref-accountable2026]{1}{]}; the responsibility
vacuum and ritual review {[}\hyperref[ref-the2026]{53}{]}, auditability
as the property that makes accountability possible
{[}\hyperref[ref-auditable2026]{12}{]}, graduated oversight
{[}\hyperref[ref-governed2026]{19}{]}, and a survey of evidence tracing
and execution provenance in LLM agents
{[}\hyperref[ref-agenttraces2026]{7}{]}, whose object is the research
literature (what tracing and attribution methods have been proposed,
organized over trace sources and trust functions) where this paper's
object is vendor documentation (what platforms emit by default); the
titles neighbor, the evidence classes do not overlap, and neither
answers the other's question. Adjacent on the practice side, an
empirical AI supply chain line measures how model releases are named,
versioned, and documented on the largest registry
{[}\hyperref[ref-ptmnaming2023]{40}, \hyperref[ref-semverhf2024]{46},
\hyperref[ref-misalignment2026]{31}{]}; it measures practice quality on
a hub where this paper grades platform record classes, and Section 6.2
takes its findings as the model-layer instance of the lineage claimed
there. Two measurements published in the weeks this survey ran are its
closest neighbors and its independent corroboration, from two different
angles. \emph{Silent Updates} {[}\hyperref[ref-silentupdates2026]{48}{]}
examines post-deployment disclosure across nine first-party API
providers and seven inference hosts and finds that ``no provider in our
sample published information allowing an external party to verify that
the artifact being served is the same one referred to in this
documentation'': the disclosure-side reading of the chain-of-custody
break whose record-side expressiveness this paper grades, over a
different sample, with neither the default/opt-in/absent protocol, the
nominal split, nor a depth measurement. And an artifact-verifiability
study over four decentralized-build package ecosystems
{[}\hyperref[ref-artifactverif2026]{9}{]} finds that provenance
attestations, where present, ``do not provide complete rebuild
specifications'': the registry-side complement of the release-surface
measurement in Section 7. Except for \emph{justifiability} (Section
1.1), terms are adopted from cited prior work at their point of use:
\emph{adherence gap} {[}\hyperref[ref-adherencegap2026]{4}{]};
\emph{residual category} for the undetermined counts
{[}\hyperref[ref-star_bowker2007]{52}{]}.

\subsection{3. The specification}\label{the-specification}

A promotion, the transition of an AI system between environments, is the
unit of analysis. You can verify a promotion; you cannot verify an
organization. A promotion is justifiable when four conditions hold, each
a property of the \emph{path} from evaluation to deployment rather than
of any single artifact. The four-way shape follows prior frameworks
{[}\hyperref[ref-prooforstop2026]{38}, \hyperref[ref-where2026]{54},
\hyperref[ref-auditable2026]{12}{]}; the definitions are quoted from the
program's frozen definitions document, whose content hash is pinned in
the OSF registration.

\textbf{Artifact continuity.} The exact artifact that was evaluated is
the artifact that was deployed. Verified by content-addressed comparison
of the built package, not by build number or stage name.

\textbf{Behavioral continuity.} The full behavioral tuple that was
evaluated is the tuple that is running: source, model version,
instructions, retrieval configuration, tool definitions, runtime
configuration, environment. Artifact continuity is necessary but not
sufficient; identical code under a different model version breaks
behavioral continuity while preserving artifact continuity.

\textbf{Evidence continuity.} The evidence cited in support of the
promotion corresponds to that behavioral tuple and was generated from it
rather than from a predecessor. It breaks when an evaluation is reused
across a change it did not observe.

\textbf{Authority continuity.} The authorization actually granted covers
the thing actually deployed, in the right scope, by an entitled actor,
in the right temporal order relative to the evidence. The temporal-order
limb is this specification's second declared residue: no surveyed or
swept framework computes the order of an approval against the evidence
it rests on as an admissibility condition on the promotion. The
condition is a formal check over the pair of records, not a measure of
review substance; whether an approval reflects consideration or ritual
review {[}\hyperref[ref-the2026]{53}{]} is a question it does not ask.

\subsubsection{3.1 The partition that makes the behavioral condition
measurable}\label{the-partition-that-makes-the-behavioral-condition-measurable}

Behavioral continuity cannot be decided by digest equality over the
whole record, because a promotion crosses environments and the
environment is \emph{supposed} to differ. The determination requires an
explicit partition, declared as constants in the instrument rather than
inferred: \textbf{invariant by contract}, where a mismatch is blocking
(the source digest, the dependency-set digest, the instruction digests,
the interface-contract digests, the runtime library versions, and the
behavior keys naming models, model location, tenant, and guardrail
policy), against \textbf{variant by design}, reported and never blocking
(the environment label, the configuration-file digest, and any behavior
key whose purpose is to differ per environment). A framework that omits
the partition either reports false breaks on every promotion or,
comparing nothing, reports none. Tuple-dependence itself is established
{[}\hyperref[ref-sculley2015]{44}, \hyperref[ref-from2026]{17}{]} and
its representation is being standardized in bill-of-materials and
reference-model form {[}\hyperref[ref-building2026]{13},
\hyperref[ref-agentriskbom2026]{6}, \hyperref[ref-acm2026]{3}{]}; every
occupant either compares whole-record equality or compares nothing. The
partition is the measurement decision, and it is this specification's
first declared residue.

\subsubsection{3.2 Nominal versus content-addressed
reference}\label{nominal-versus-content-addressed-reference}

Cross-cutting all four conditions, a reference is either
\textbf{content-addressed} (a digest of the state itself, which cannot
drift without the state changing) or \textbf{nominal} (a label
maintained by convention, which can drift independently of what it
names). The distinction is established at specification level and
measured at scale for artifact identity {[}\hyperref[ref-slsa2025]{50},
\hyperref[ref-mutating2026]{32}{]}, where expectations bind to names,
provenance binds to digests, and names drift; the same
claimed-versus-attested split has now been measured at the commit layer
across billions of commits {[}\hyperref[ref-claimedattested2026]{14}{]}.
It is carried here because it is where the survey's central finding
lives: a system can be sound in one layer and broken in the other
simultaneously (the motivating estate held a manifest whose every
content-addressed field was correct while its nominal release identity
had been wrong for twelve consecutive releases), and a single
traceable-or-not verdict cannot express the state.

\subsubsection{3.3 The default definition, and the
rubric}\label{the-default-definition-and-the-rubric}

Two operational definitions govern everything measured below.

\textbf{Default.} A datum is emitted \emph{by default} when the platform
records it as part of its standard mechanism for that record class,
given only that the mechanism is used at all (an artifact is uploaded,
an approval is requested, an agent is deployed); it is \emph{opt-in}
when recording requires enabling a distinct feature (an additional
action, an environment flag, a plugin); it is \emph{absent}, always as a
bounded absence, when the consulted documentation pages, on point for
the dimension and scanned with tuple-capable search terms, describe no
such capability. Declaring where an artifact lives is using the
mechanism, not opting in.

\textbf{Depth.} Every promotion path gets an ordinal assurance depth,
from the frozen definitions document, twice: \textbf{intended} from its
pipeline definition and \textbf{realized} from its published evidence,
where a binding counts only if populated with a content-addressed value.
The scale carries no arithmetic; no magnitude claim attaches to rung
differences.

{\def\LTcaptype{none} 
\begin{longtable}[]{@{}
  >{\raggedright\arraybackslash}p{0.48\linewidth}
  >{\raggedright\arraybackslash}p{0.48\linewidth}@{}}
\toprule\noalign{}
\begin{minipage}[b]{\linewidth}\raggedright
Depth
\end{minipage} & \begin{minipage}[b]{\linewidth}\raggedright
Definition
\end{minipage} \\
\midrule\noalign{}
\endhead
\bottomrule\noalign{}
\endlastfoot
0 & Transition occurs with no gate \\
1 & Authority gate only: an entitled actor approves, no evidence
required \\
2 & Evidence gate: an automated verification must pass, unbound to
artifact identity \\
3 & Bound evidence gate: verification passes and its result is bound to
a content-addressed identity \\
4 & Bound evidence gate plus authority, where the authorization refers
to the bound identity \\
\end{longtable}
}

One reading rule keeps the scale honest against the specification's own
dissociation argument. Rungs 1 and 2 are not ordered by assurance: an
authority gate and an unbound evidence gate assure different conditions,
and the conditions come apart (Section 3), so the rubric reads the
\emph{evidence side} of a pipeline: a pipeline with both an approval
gate and an unbound check sits at rung 2, its authority gate absorbed,
and authority re-enters the scale only at rung 4, where the
authorization refers to the bound identity. The ordinal is therefore
valid within a pipeline, comparing realized depth against that same
pipeline's declaration (a declared evidence gate whose exhaust shows
only authority is a shortfall of the declared class, whatever one thinks
of the 1-versus-2 ordering), and every cross-pipeline claim in Section 7
uses only the distinction between the binding rungs (3 and 4) and the
rest, never the ordering of 1 against 2.

\subsection{4. The instrument, in its public
form}\label{the-instrument-in-its-public-form}

The instrument is a set of small read-only checkers over artifacts a
platform already publishes. No new system of record, no instrumentation
added to any pipeline, no producer signatures. This is the inverse of
the supply chain frameworks' deployment model: in-toto and SLSA place
obligations on producers and verify what cooperating producers emitted
{[}\hyperref[ref-intoto2019]{23}, \hyperref[ref-slsa2023]{49}{]}; here
the producers are not parties to the measurement, because that is the
situation nearly every organization deploying agents is in today. The
property that makes the instrument worth a paper is that the comparison
it computes, realized depth against declared depth, requires asking no
one: it reads configuration for the declaration and published exhaust
for the realization, from different sources by construction, so the gap
cannot be an artifact of one record describing itself.

Three design rules. \textbf{Blind to outcomes:} the depth classifier
reads pipeline definitions and policy configurations only, never defect
data or gate pass rates; this preserves the registered second-site
study. \textbf{Declared query bounds:} platform APIs truncate silently,
so every count states its frame and absence claims state their bounds.
\textbf{The residual is counted, not discarded:} everything the evidence
cannot support is graded undetermined and reported, never dropped; an
instrument that reports only what it could decide overstates what
artifacts can do {[}\hyperref[ref-sbom_noedges2026]{43},
\hyperref[ref-star_bowker2007]{52}{]}.

\subsubsection{4.1 The public depth study}\label{the-public-depth-study}

\textbf{Frame, frozen before collection.} 30 public GitHub repositories
with release pipelines on GitHub Actions, in two declared strata of 15:
stratum A, repositories documented as adopting artifact attestation or
SLSA tooling (the ceiling stratum, where declared depth should be
highest); stratum B, high-activity release repositories from a
stars-ranked frame. Queries and candidate lists were recorded verbatim
before truncation; the frame froze before collection began, and no
repository was added or dropped after first read. Two repositories used
to dry-run the collector were excluded from the frame before it froze.

\textbf{Collection, deterministic and archived.} A committed collector
retrieved, for every frame repository, its workflow definitions,
rulesets, environments, releases with assets, per-asset attestation
lookups, and deployment records: 1,116 raw API responses archived under
a per-repository manifest with a SHA-256 hash for every file. The
archive is the study's entire evidence base; neither grading pass
fetched anything live. Unavailability is data: the branch-protection
endpoint is admin-only and returned not-readable on all 30 repositories,
so intended depth is graded from rulesets, environments, and workflow
definitions only; and the attestation surface splits three ways (7
repositories with a found attestation, 6 where lookups returned none,
and 17 where release assets exposed no digests so the lookup was never
attempted), a split the grading protocol forbids collapsing, because a
returned \emph{none} is a result while \emph{not attempted} is an
absence of measurement.

The collector was dry-run on two repositories excluded from the frame
before it froze, and the dry run's lessons were recorded as binding
reading rules rather than silently applied: empty ruleset listings are
ambiguous (the API returns the same empty list for ``no rulesets'' and
``none readable'') and support no depth determination; a repository's
attestation status is per asset, not per repository, because one
repository returned a found attestation for its platform binaries and
none for its checksum file; and one collector defect discovered during
the run (environment names containing slashes break the detail
endpoint's URL and return spurious not-found responses) was verified
read-only against the live API, recorded with its blast radius, and
deliberately not patched mid-study, since changing what the collector
collects between freeze and grading would have made the archive
unreproducible from the committed code.

\textbf{Grading, twice, blind.} Two independent passes applied the
printed rubric to the archive, the second without access to the first. A
binding counted only if populated with a content-addressed value; an
attestation counted only if actually found for the release assets; an
approval counted only if a record of it exists in the archive.
Cell-level agreement was 23 of 30 on intended depth and 19 of 30 on
realized; the raw figures are reported because they bound what the
reconciliation can claim, and the stronger reliability fact is
distributional: both passes, before any reconciliation, independently
identified the same five stratum-A repositories as full realizations, so
the divergent cells dispute which lower rung, never whether the binding
was realized. All 18 divergences were then resolved from the archive
with the resolution basis recorded per cell, and a committed
deterministic analyzer recomputes the final distributions from the
graded files.

\textbf{One correction, dated and archived.} After reconciliation, an
external check of the 17 repositories whose release assets exposed no
digests established from the archive that all 17 publish
\textbf{source-only releases}: zero uploaded assets, only the
auto-generated source archives, which the platform never digests. The
built artifacts of these pipelines publish to registries outside the
frame, so the release surface can neither show nor refute the binding
their workflows declare. This forced a grading rule the original
protocol lacked, applied as a dated correction with the reconciled file
preserved unmodified: where the graded release carries no uploaded
assets, realized depth at the identity-binding rungs is undetermined,
never a measured rung. The rule separates ``no digest because the
platform computed none for this release'' from ``no digest for the
assets this pipeline publishes,'' and it is the same discipline already
applied to the admin-only branch-protection surface: unavailability is
data, not failure. Nine cells moved from a measured rung to
undetermined; the corrected file, the rule, and the per-cell bases are
in the archive.

\textbf{Origin note.} The instrument was first built as an audit tool
over one operating estate's exhaust, which is where the motivating case
of Section 1 was detected; that estate had built itself a bespoke
content-addressed binding artifact, which is exactly what no surveyed
platform emits by default. The public form reported here uses public
material only, and the program's estate-facing evidence routes to its
other papers.

\subsection{5. The surveys}\label{the-surveys}

\textbf{Protocol.} One fixed three-label protocol (Section 3.3's default
definition) over four dimensions per platform: artifact digest (does the
standard mechanism record a content digest of the deployable or model
artifact), approval record (actor identity and timestamp on
deployment-approval events), provenance emission (provenance, lineage,
or attestation records emitted for builds, models, or agent
configurations), and behavioral-tuple identity (a content-addressed
identity over the tuple of Section 1.1). Every tuple cell additionally
grades \textbf{nominal versioning} separately: does the platform record
the tuple, or most of it, by default as an immutable named or numbered
version? A version integer behind a mutable pointer is nominal; a digest
over configuration is content-addressed. Every default or opt-in grade
carries a verbatim quote, URL, and fetch date; every absent grade names
the consulted pages and is falsifiable by naming a page; tuple cells are
graded only after searches including the tuple-capable terms (version,
snapshot, immutable, digest, sha256, checksum, lineage, provenance,
system prompt, instructions, tool definition).

\textbf{Pinning.} Every consulted page was pinned at grading time: the
retrieved bytes hashed, the hash and fetch date appended to an
append-only pin log (779 records at sealing), and a public archive
snapshot requested. Where the public archiver declined the URL, the
retrieval hash is the durable evidence; the log records which. A re-run
of the survey is therefore a true replication: it can determine whether
a page changed, not merely disagree about what it says.

\textbf{Two passes, the second blind.} Every cell in both classes was
graded twice by independent sessions, the second working from the
protocol text alone, without access to the first pass's grades, the
survey record, or the pin log. Raw agreement: the original five CI/CD
platforms, 17 of 20 cells; the fifteen-platform CI/CD expansion, 55 of
60, with all five divergences in the approval dimension; the 27
agent-class platforms, 78 of 108 on main grades and 21 of 27 on the
nominal split. The figures are raw agreement over a three-label space
skewed toward absent and default; no chance-corrected statistic is
reported at these cell counts. Every divergence was resolved from the
better-evidenced source, re-fetching the live page where the two quotes
conflicted, and every resolution is recorded with its basis in the
survey record; the sealed tables below are the reconciled ones, and the
raw pass files are archived unmodified.

The divergences themselves are method data, and their shapes recur.
Vocabulary misses: one pass graded Bitbucket's approval record absent
because it searched for the word ``approval'', which Bitbucket does not
use; the platform's unblock mechanism is the manual trigger step, and
the deployment summary records who triggered it and when, so the cell
resolved to default on the first pass's evidence. Criterion application:
both passes found GCP Cloud Build's approval object with identical
quotes and graded it differently; under the fixed criterion, configuring
an approval-requiring trigger is requesting approval, which is using the
mechanism, so the record that follows automatically is default. Sharper
distinctions won in either direction: Semaphore resolved to opt-in
because the actor-bearing record exists only under a configured feature,
and a bare manual promotion records no actor. In the agent class the
modal divergence ran one way: the first pass was systematically more
generous than the blind pass, and most divergent cells resolved toward
the stricter reading, including every divergence on the tuple dimension.
One interpretive choice moved cells and is recorded as a standing rule
rather than silently absorbed: cloud audit logging (management-event
trails that record that an API call occurred) is not an approval record,
because it captures the act without the approval semantics; where a
platform's only actor-and-timestamp trail is its audit log, the approval
cell is graded absent and the audit log is carried in the cell's notes
as a named near-miss.

\subsection{6. Results I: what the default record can
express}\label{results-i-what-the-default-record-can-express}

\subsubsection{6.1 The CI/CD class, 20
platforms}\label{the-cicd-class-20-platforms}

The original five platforms, graded first and carried as the detail
exemplar (every cell leads with its label; anything after the
parenthesis is a gloss, not a grade):

{\def\LTcaptype{none} 
\begin{longtable}[]{@{}
  >{\raggedright\arraybackslash}p{0.192\linewidth}
  >{\raggedright\arraybackslash}p{0.192\linewidth}
  >{\raggedright\arraybackslash}p{0.192\linewidth}
  >{\raggedright\arraybackslash}p{0.192\linewidth}
  >{\raggedright\arraybackslash}p{0.192\linewidth}@{}}
\toprule\noalign{}
\begin{minipage}[b]{\linewidth}\raggedright
Platform
\end{minipage} & \begin{minipage}[b]{\linewidth}\raggedright
Artifact digest
\end{minipage} & \begin{minipage}[b]{\linewidth}\raggedright
Approval record (actor, time)
\end{minipage} & \begin{minipage}[b]{\linewidth}\raggedright
Provenance emission
\end{minipage} & \begin{minipage}[b]{\linewidth}\raggedright
Tuple identity
\end{minipage} \\
\midrule\noalign{}
\endhead
\bottomrule\noalign{}
\endlastfoot
GitHub Actions (github.com) & default (\texttt{digest} on every v4+
upload; not on GHES; covers the upload archive) & default (approvals
API: approver, state, comment; timestamp default in the audit-log
record) & opt-in (attest action) & absent \\
GitLab CI & absent (no digest on the default record; sha256 only inside
the opt-in provenance metadata) & default (approver, status,
\texttt{created\_at} on the deployment record) & opt-in (runner flag) &
absent \\
Jenkins (core) & opt-in (fingerprints, MD5; Maven jobs auto) & opt-in
(\texttt{submitterParameter} captures actor; no timestamp documented) &
absent (no first-party emission; in-toto via community plugin) &
absent \\
Azure DevOps & absent (no digest field on the artifact objects) &
default (\texttt{actualApprover}, \texttt{initiatedOn},
\texttt{lastModifiedOn}, history) & absent (consulted first-party docs
describe none) & absent \\
Buildkite & default (SHA-1 unconditional; SHA-256 by agent version) &
default (\texttt{unblocked\_by}, \texttt{unblocked\_at}) & opt-in
(first-party plugin) & absent \\
\end{longtable}
}

The expansion added CircleCI, Travis CI, TeamCity, Bamboo, Drone,
Woodpecker, Argo, Tekton, Spinnaker, Harness, AWS
CodePipeline+CodeBuild, GCP Cloud Build, Bitbucket Pipelines, Semaphore,
and Codefresh under the identical protocol. The sealed class counts, all
20 platforms:

{\def\LTcaptype{none} 
\begin{longtable}[]{@{}llll@{}}
\toprule\noalign{}
Dimension (CI/CD, n=20) & default & opt-in & absent \\
\midrule\noalign{}
\endhead
\bottomrule\noalign{}
\endlastfoot
artifact digest & 5 & 4 & 11 \\
approval record & 13 & 2 & 5 \\
provenance emission & 0 & 6 & 14 \\
behavioral-tuple identity & 0 & 0 & 20 \\
\end{longtable}
}

Three facts carry the table. The approval record is the class's most
uniformly default rung, and its best exemplar is complete: GCP Cloud
Build's ApprovalResult carries approver account, approval time,
decision, and comment on every approval-gated build. Provenance emission
is default nowhere in the class: where it exists at all it is an action
you add, a runner flag, an installable subsystem (Tekton Chains), or a
step template (Harness), never the standard mechanism's own record. And
the behavioral tuple does not appear in the class's record vocabulary at
all: 20 bounded absences, which is the expected shape (CI/CD platforms
move artifacts, not behavioral configurations) and exactly why the
declared depth of Section 7.1's public pipelines tops out at the
artifact-bound rung.

\subsubsection{6.2 The agent and model-serving class, 27
platforms}\label{the-agent-and-model-serving-class-27-platforms}

Amazon Bedrock, Microsoft Foundry, Google Vertex AI, OpenAI platform,
Anthropic Claude platform, Ollama, Hugging Face, Azure ML, SageMaker,
Weights \& Biases, MLflow, LangSmith/LangGraph, Databricks, Modal,
Baseten, Together AI, Anyscale, Replicate, Fireworks, Groq, Cohere,
Mistral, Dify, CrewAI, Vercel AI, RunPod, and NVIDIA NIM/NGC. The sealed
class counts:

{\def\LTcaptype{none} 
\begin{longtable}[]{@{}llll@{}}
\toprule\noalign{}
Dimension (agent/ML, n=27) & default & opt-in & absent \\
\midrule\noalign{}
\endhead
\bottomrule\noalign{}
\endlastfoot
artifact digest & 7 & 2 & 18 \\
approval record & 8 & 5 & 14 \\
provenance emission & 11 & 3 & 13 \\
behavioral-tuple identity (content-addressed) & \textbf{0} & 2 & 25 \\
nominal tuple versioning (the split, same cells) & 16 & 2 & 9 \\
\end{longtable}
}

\begin{figure}
\centering
\pandocbounded{\includegraphics[keepaspectratio,alt={What the default record can express, per dimension, both classes: sealed reconciled counts. Content-addressed tuple identity is default nowhere; nominal versioning is the agent class's arriving default answer.}]{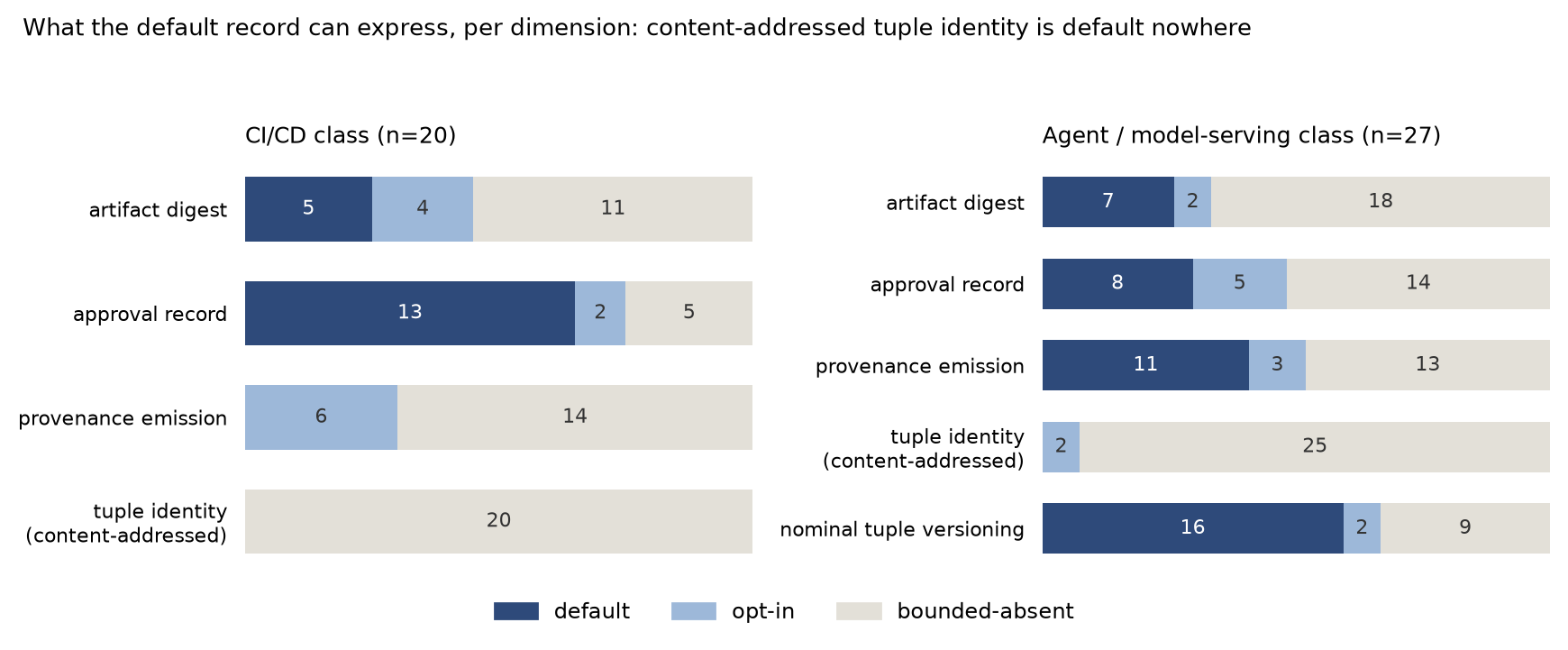}}
\caption{What the default record can express, per dimension, both
classes: sealed reconciled counts. Content-addressed tuple identity is
default nowhere; nominal versioning is the agent class's arriving
default answer.}
\end{figure}

The bottom two rows are the survey's center, stated at the strength a
graded survey can support: \textbf{across the 47 platforms, in a survey
of 188 double-graded cells, we found no platform, in either class, whose
default record emits a content-addressed identity over the behavioral
tuple}, including all 27 platforms in the class where it would most
plausibly exist. The claim is a bound, not an absolute: it is robust to
the discovery of a misgraded cell in a way a bare zero would not be, and
the falsifier is designed to refine it rather than destroy it, since
every absence names its consulted pages and is answered by naming a
page. The bound's reliability is reported at the column that carries it,
not only in aggregate. Raw agreement on the tuple-identity column was 23
of 27 in this class and 20 of 20 in the CI/CD class, and all four
divergences ran in one direction, against the paper's own claim: the
non-blind first pass, the pass that knew the hypothesis, graded default,
the one label that would refute the headline, where the blind pass
graded opt-in or absent. A grader captured by the hypothesis errs the
other way, so the divergence direction is itself evidence against
capture, and every resolution moved away from default under the
partial-cover instruction, not toward the result the paper needs but
toward the stricter reading of the criterion. In no tuple cell, in
either class, did both passes agree on default; the blind pass alone,
unreconciled, grades the column default on zero of 47 platforms. The
aggregate agreement figures of Section 5 are therefore mostly dispersion
in the approval and provenance columns (16 and 17 of 27 in this class),
where the gated-feature ambiguities live, and not in the column the
headline rests on. The two opt-in cells are partial even then. The
near-misses are named rather than counted as grades: Ollama's manifest
digest covers weights, template, system prompt, and parameters but not
the serving configuration around them; Hugging Face's git revision
content-addresses a repository, and its serving endpoints default to the
floating branch head rather than the pin. Meanwhile \textbf{immutable
nominal versioning of the tuple is the class's arriving default answer,
on 16 of 27 platforms}: version integers or named revisions behind
mutable pointers, with explicit immutability language (Bedrock, Foundry,
Azure ML, SageMaker) or version-pinning APIs (Anthropic's managed agents
version the full tuple of model, instructions, tools, and skills; the
version is a counter, not a digest). The class is institutionalizing a
layer whose failure is already measured one level down: on the largest
model registry, name-based versioning left 40.87 percent of model-weight
changes unrepresented in either the name or the documentation
{[}\hyperref[ref-semverhf2024]{46}{]}, naming practice itself is
inconsistent enough to mislead reuse
{[}\hyperref[ref-ptmnaming2023]{40}{]}, and producers and consumers of
the same models disagree about where critical metadata should be
recorded at all {[}\hyperref[ref-misalignment2026]{31}{]}.

The gloss layer of the sealed cells records how nominal the nominal
layer is. Vertex AI's agent revisions are ``always enabled'' and
immutable by documentation, but the versioned fields hold storage URIs,
so a revision pins pointers rather than bytes, and fields outside the
versioned set mutate across all revisions. Vercel's deployments are
immutable, while team-level gateway routing rules can rewrite which
model actually serves without any redeployment, drifting the tuple under
an unchanged deployment identity. Mistral's document library versions
are ``immutable'' by stated policy, with no digest to check the policy
against. Dify's publish history permits deleting old versions. And where
provenance emission \emph{is} default in this class (SageMaker, Vertex,
Azure ML lineage), the binding is referential, by storage URI rather
than by hash, which SageMaker's own documentation states directly;
NVIDIA NIM/NGC is the survey's one default provenance emission with
content-addressed reach (signed catalog models with per-digest SBOM),
scoped to NVIDIA-published artifacts. The two largest model vendors are
moving in opposite directions on exactly this paper's question: OpenAI
is retiring server-side tuple versioning outright (its prompt-object and
agent-builder surfaces are scheduled to shut down 2026-11-30, with the
record routed to the customer's own repository), while Anthropic is
introducing it; a survey re-run after that date should expect OpenAI's
nominal cell to fall to absent.

The approval dimension tells the class's governance story in miniature.
Where CI/CD platforms treat deployment approval as a first-class record
(Section 6.1), most agent platforms either have no approval record class
at all (14 of 27 absent, typically because the deploy act itself is the
only gate) or gate the record behind administration features: the
class's most complete deployment-event record, carrying a typed actor
and timestamp, ships toggled off by default and is unrecoverable for the
period it was off, which under the protocol is opt-in however complete
the record is once enabled. The same asymmetry appears inside single
platforms: several record who \emph{requested} a deployment but never
who approved it, and one lineage system overwrites its last-modified-by
field on any later edit, so the approval-shaped datum it holds is the
most recent editor, not the authorizing actor. For the authority
condition of Section 3 this means the temporal-order check that is
cheaply computable across most of the CI/CD class is computable on only
a minority of the agent class, and the deployment acts most in need of
an authority record are the ones least likely to have one by default.

\subsubsection{6.3 The ladder}\label{the-ladder}

Read as one structure, the two classes say what the four conditions of
Section 3 can reach from default records, without bespoke
instrumentation. Artifact continuity is determinable unevenly (digests
default on a quarter of the CI/CD class and a quarter of the agent
class). Authority continuity's approval rung is the most uniformly
default record in the survey; its temporal order is computable wherever
the approval and evidence records carry timestamps, and no platform or
framework computes it. Evidence continuity is determinable only through
a binding artifact the estate builds itself: evidence schemas carry
optional identity fields {[}\hyperref[ref-intotospec2024]{24}{]}, and
nothing in the default record populates them. Behavioral continuity
splits on Section 3.2's distinction, and the split is the finding: not
recorded at all in the CI/CD class; recorded by default in the agent
class as immutable \emph{versions}, so the condition is determinable
there \textbf{nominally}, against integers behind mutable pointers, and
content-addressed nowhere. A nominal determination inherits the nominal
layer's failure mode. The digests themselves are a syntactic proxy
(records with identical digests can diverge through channels the record
does not carry), which is a stated limit of the instrument, not of the
survey.

\subsection{7. Results II: the measured gap, and why it is
structural}\label{results-ii-the-measured-gap-and-why-it-is-structural}

\subsubsection{7.1 The public depth-gap
distribution}\label{the-public-depth-gap-distribution}

The corrected per-repository grades, both strata (the figure lists every
repository; undetermined is a counted outcome, and the two blind passes'
own realized distributions are shown beside the adjudicated one, because
a reader must be able to see what the reconciliation did):

{\def\LTcaptype{none} 
\begin{longtable}[]{@{}lllllll@{}}
\toprule\noalign{}
Intended & 0 & 1 & 2 & 3 & 4 & undet. \\
\midrule\noalign{}
\endhead
\bottomrule\noalign{}
\endlastfoot
Stratum A (n=15) & 0 & 0 & 0 & 15 & 0 & 0 \\
Stratum B (n=15) & 0 & 5 & 6 & 4 & 0 & 0 \\
\end{longtable}
}

{\def\LTcaptype{none} 
\begin{longtable}[]{@{}lllllll@{}}
\toprule\noalign{}
Realized, stratum A & 0 & 1 & 2 & 3 & 4 & undet. \\
\midrule\noalign{}
\endhead
\bottomrule\noalign{}
\endlastfoot
pass 1 (blind) & 0 & 8 & 0 & 5 & 0 & 2 \\
pass 2 (blind) & 0 & 6 & 4 & 5 & 0 & 0 \\
adjudicated + corrected & 0 & 0 & 2 & 5 & 0 & 8 \\
\end{longtable}
}

{\def\LTcaptype{none} 
\begin{longtable}[]{@{}lllllll@{}}
\toprule\noalign{}
Realized, stratum B & 0 & 1 & 2 & 3 & 4 & undet. \\
\midrule\noalign{}
\endhead
\bottomrule\noalign{}
\endlastfoot
adjudicated + corrected & 0 & 9 & 2 & 2 & 0 & 2 \\
\end{longtable}
}

One honesty note before the readings: stratum A's intended column is the
sampling frame restated, not a result. The stratum was selected for
documented attestation adoption, so ``15 of 15 declare rung 3'' verifies
the frame rather than discovering anything; what is measured is the
realized row against it. The reliability of that row is carried by the
pass-level table: both blind passes independently found the same five
repositories at rung 3, so the 18 cell-level divergences (both passes'
full grids are archived) dispute which lower rung, never whether the
binding was realized. The corrected row must be read with its provenance
stated plainly: it is not what the passes converged on. Pass 1 graded
two stratum-A cells undetermined and pass 2 none; the corrected row
carries eight there, and two in stratum B, because the source-only rule
of Section 4.1 is a post hoc rule change, adopted after reconciliation
on external review and applied uniformly to every affected cell of both
strata, not an output of the passes' disagreement. Stratum B's two
undetermined are the same rule: two repositories whose graded releases
carry no uploaded assets.

\begin{figure}
\centering
\pandocbounded{\includegraphics[keepaspectratio,alt={Declared versus realized assurance depth for all 30 repositories. Light dots mark intended depth from workflow definitions; dark dots mark realized depth from published exhaust; arrows point from declared to realized; source-only rows are undetermined by rule.}]{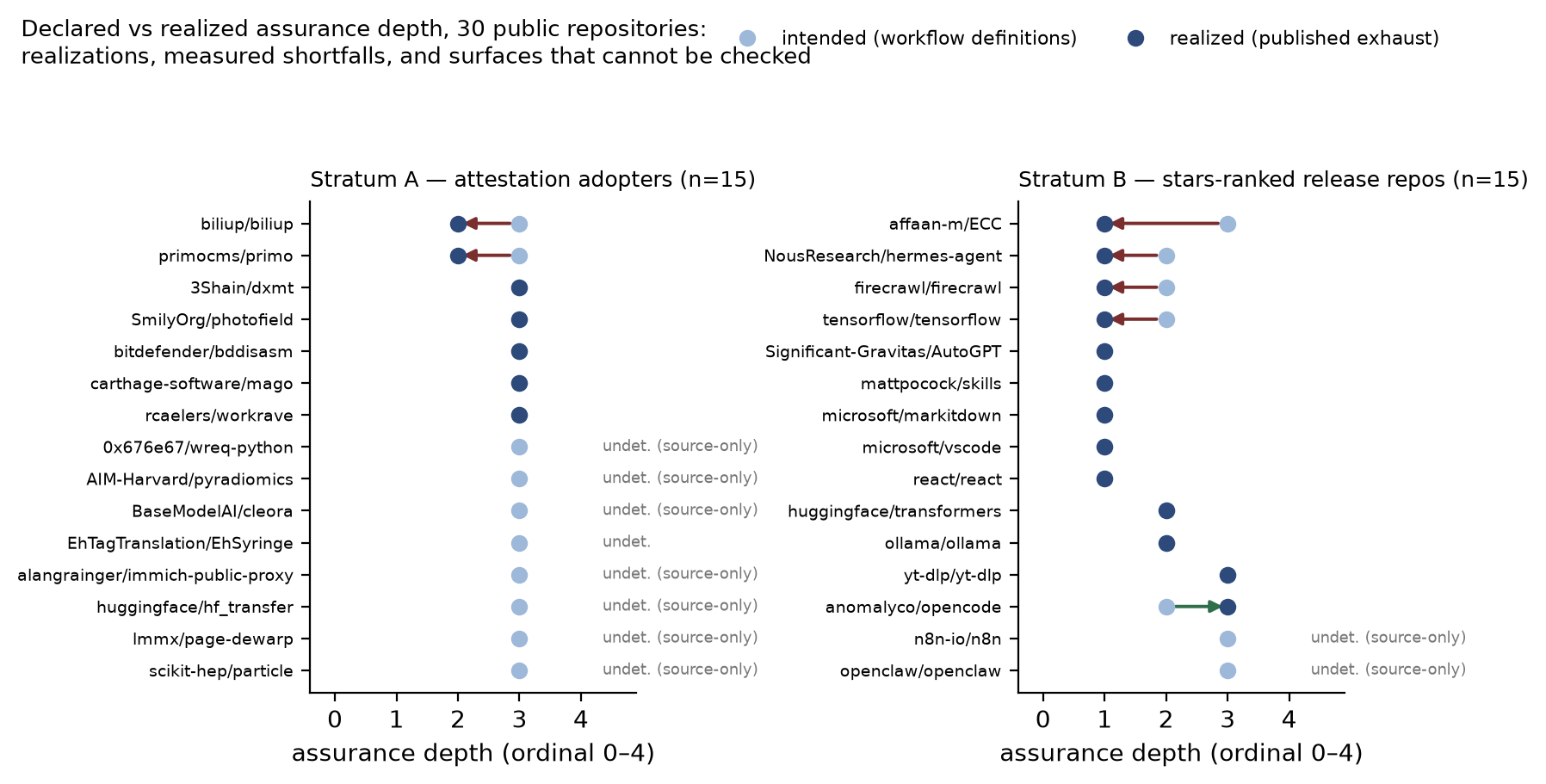}}
\caption{Declared versus realized assurance depth for all 30
repositories. Light dots mark intended depth from workflow definitions;
dark dots mark realized depth from published exhaust; arrows point from
declared to realized; source-only rows are undetermined by rule.}
\end{figure}

Four readings, in order of weight. \textbf{The sharpest finding is a
verifiability hole: seven of the fifteen attestation adopters publish
source-only releases, so the binding their workflows declare has no
artifact surface on the platform where they declare it.} The declared
attest machinery operates on artifacts the release never carries; those
artifacts live on package registries that are themselves public, and the
frame did not examine them, so the accurate statement is precise: the
binding is uncheckable on the surface where the declaration appears, and
whether it holds on the registry is unknown because the study did not
look. The declaration and the verifiable evidence, where it exists, live
on different public surfaces with nothing in the exhaust linking them;
the registry-side half of that seam is exactly what release-authority
and artifact-verifiability measurement over package ecosystems has begun
to read {[}\hyperref[ref-on2026]{35},
\hyperref[ref-artifactverif2026]{9}{]}, which makes the cross-surface
join the natural follow-up frame for this instrument. \textbf{Two
realization rates, and both are needed.} Five of the fifteen adopters
realize the declared binding end to end (one third of the stratum):
per-asset digests on the release, attestation lookups returning found
bundles for those digests, the chain checkable by any unauthenticated
reader. But only seven of the fifteen have a release surface that can be
measured at all, and against that denominator the same five are five of
seven. Among the adopters you can check, most check out; the finding is
that most cannot be checked on the surface where they declare.
\textbf{The instrument returns zero as readily as it returns a gap, and
it can run negative.} Stratum B's modest declarations are mostly
realized as declared or one rung short, and one repository's published
exhaust (complete per-asset digests and found attestations) exceeds what
its workflows declare as a gate: evidence without a declared consumer,
the mirror image of a declaration without evidence. \textbf{Both
measured stratum-A shortfalls fall at the binding rung, an observation
of n = 2}: two adopters publish digest-bearing assets for which no
attestation is discoverable, realized rung 2 against declared rung 3.
That is consistent with the prediction that the gap concentrates at the
rungs requiring identity binding, and far too small to establish it;
stratum B's measured shortfalls, for contrast, sit mostly at the
unbound-evidence rung.

Every realized grade in both strata is a lower bound, stated once for
the whole table: the admin-only branch-protection surface was excluded
by rule on all 30 repositories, and ambiguous empty ruleset listings
support no determination, both of which can only depress declared and
realized depth as graded; a gap measured under those exclusions survives
them.

The distribution also does work the specification needs. Section 3's
four-way decomposition would be over-engineered if failures were
monolithic; this table is the conditions coming apart at population
scale. Stratum A's measured cells separate an evidence shortfall
(verification present, binding absent: the two rung-2 adopters) from
full realization; the negative-gap repository separates published
evidence from declared authority (evidence flowing with no gate
consuming it); the source-only majority separates declared machinery
from any public surface that could carry its evidence; and the authority
condition separates on the archive rather than on the ordinal. Because
the rubric absorbs authority below rung 4 (Section 3.3's reading rule),
a per-repository declared-authority flag was derived deterministically
from the archived rulesets and environments
(\texttt{authority-flags.json}): five of stratum B's six rung-2
declarations also declare an approval gate, so those pipelines declare
evidence and authority together and realize only authority, while its
five rung-1 pipelines declare authority and no evidence condition at
all. No single traceable-or-not verdict expresses any row of that table,
which is the operational argument for the decomposition, made here by a
population rather than by one estate.

Two boundary facts locate this study on the paper's ladder, and they
pull in opposite directions, which is why both are stated. Downward:
behavioral depth is structurally out of reach in this population, since
these are all CI/CD pipelines whose record vocabulary contains no
behavioral tuple (Section 6.1's fourth row), the survey's ceiling met in
the wild. Upward: at the artifact layer this population is the
\emph{favorable} case, because the platform it runs on is one where this
paper's own survey grades artifact digests as default. Inside this frame
the record class the declared depth needs exists and is default, so
nothing measured here is an instance of Section 7.2's documentary
finding, and the paper does not claim it is. The seven source-only
repositories are not short a record class: they publish no uploaded
assets, so there is nothing for the default mechanism to digest: a
publishing-surface choice, not a record-expressiveness limit, which is
why the verifiability hole stands as this paper's third result rather
than as the documentary finding's exhibit. What the floor measurement
shows is that even under default-available records, realization of
declared artifact-layer assurance is partial, fragmented across
surfaces, and measurable without asking anyone.

\subsubsection{7.2 The documentary finding, its corollary, and its
clock}\label{the-documentary-finding-its-corollary-and-its-clock}

\begin{quote}
\textbf{Documentary finding.} Across 47 platforms in two classes, graded
twice under a fixed protocol with the second pass blind, no consulted
documentation describes default emission of a content-addressed identity
of the behavioral tuple. The agent class now records the tuple by
default as immutable \emph{versions} on 16 of 27 platforms, making
behavioral continuity determinable there nominally but not at the
content layer; the features that approach content addressing are opt-in,
partial, or defeated by mutable indirection the record does not carry.
\end{quote}

\begin{quote}
\textbf{Corollary (definitional, conceded as such).} A pipeline whose
declared assurance depends on an identity its records do not carry
cannot realize its declared depth. This follows from what ``realize''
means and earns no empirical credit; its work is to locate where the
documentary finding bites: on such pipelines, an adherence gap at the
identity-binding rungs is not organizational carelessness but the
record's expressiveness meeting a declaration it cannot carry. The
corollary's class is behavioral-rung pipelines without a bespoke binding
artifact. Section 7.1's population is deliberately \emph{not} in that
class (its needed records are default on its platform), which is exactly
why the floor it measures is informative: partial realization there is
attributable to practice and surface fragmentation, never to the record,
so the two results bound the problem from opposite sides rather than
illustrating each other.
\end{quote}

\begin{quote}
\textbf{Prevalence conjecture (untested at the behavioral rung).}
Pipelines whose declared assurance depends on behavioral identity, with
no bespoke binding artifact, are the common case, so the gap will be
observed widely there as tuple-dependent workloads reach delivery
pipelines. The artifact-layer analogue is no longer conjecture (Section
7.1 measures it); the behavioral-rung form remains untested because the
records it needs do not exist to be graded, which is the finding.
\end{quote}

Stated as the lineage it belongs to rather than as a fresh discovery:
the nominal-versus-content-addressed problem is the one the supply chain
community has spent a decade working for artifact identity, where
expectations bind to names while provenance binds to digests and names
drift at scale {[}\hyperref[ref-slsa2025]{50},
\hyperref[ref-mutating2026]{32}{]}, and the AI supply chain literature
has since measured the same failure at the model layer, where name-based
versioning does not track the bytes and registry metadata defeats
lineage tracing {[}\hyperref[ref-semverhf2024]{46},
\hyperref[ref-ptmnaming2023]{40},
\hyperref[ref-misalignment2026]{31}{]}. The documentary finding is that
problem reappearing intact one layer further up, at the behavioral
tuple, and the platforms now beginning to record the tuple are resolving
it nominally a third time: version integers behind mutable pointers, the
layer already found insufficient twice below.

The finding is falsified by exhibiting a platform that emits a
content-addressed tuple identity by default, judged by Section 3.3's
definition; it \textbf{expires}, rather than fails, when the
standardization efforts converging on tuple representation
{[}\hyperref[ref-building2026]{13}, \hyperref[ref-agentriskbom2026]{6},
\hyperref[ref-acm2026]{3}{]} ship as defaults on a named major platform.
The clock is running: a successor SPDX release candidate exists, one
platform already content-addresses a partial tuple slice by default, the
vendor directions of Section 6.2 are moving in the survey's observation
window, and the disclosure-side measurement of the same break is now
independently on record {[}\hyperref[ref-silentupdates2026]{48}{]}, so
the pressure toward emission is arriving from two directions at once.
The surveys are re-run at this paper's first venue revision and no later
than 2027-02, under the same protocol, against the same pinned baseline,
and the finding's status is restated then.

\subsubsection{7.3 What this paper has shown, per
condition}\label{what-this-paper-has-shown-per-condition}

The specification names four conditions; the evidence does not reach
them equally, and the reader should not have to assemble the ledger from
four sections. Per condition: what the survey says the default record
can express, and what the depth study measured.

{\def\LTcaptype{none} 
\begin{longtable}[]{@{}
  >{\raggedright\arraybackslash}p{0.32\linewidth}
  >{\raggedright\arraybackslash}p{0.32\linewidth}
  >{\raggedright\arraybackslash}p{0.32\linewidth}@{}}
\toprule\noalign{}
\begin{minipage}[b]{\linewidth}\raggedright
Condition
\end{minipage} & \begin{minipage}[b]{\linewidth}\raggedright
Survey: default-record reachability
\end{minipage} & \begin{minipage}[b]{\linewidth}\raggedright
Depth study: measured?
\end{minipage} \\
\midrule\noalign{}
\endhead
\bottomrule\noalign{}
\endlastfoot
Artifact & digests default on 5 of 20 CI/CD and 7 of 27 agent platforms
& \textbf{Yes, both directions}: realized bindings (5 adopters end to
end), measured shortfalls (2, at the binding rung), and the surface
limits, all from exhaust \\
Authority & approval records the survey's most uniformly default rung
(13 of 20; 8 of 27) & \textbf{Partially}: approvals counted where
records exist; the temporal-order limb, this specification's declared
residue, is computed nowhere in this paper \\
Evidence & evidence schemas carry optional identity fields; nothing
default populates them & \textbf{No}: rung 3 checks \emph{binding} to a
content-addressed identity, which is a different property from freshness
(evidence generated from the tuple rather than a predecessor); freshness
is unmeasured here \\
Behavioral & content-addressed identity default nowhere; nominal
versioning default on 16 of 27 agent platforms & \textbf{No, and cannot
be}: the population's record vocabulary contains no tuple (Section 6.1),
which is the survey's ceiling met in the wild \\
\end{longtable}
}

Stated plainly: this paper demonstrates the artifact condition measured,
the authority condition partially measured, and the evidence and
behavioral conditions specified with their measurement obligations
located (the evidence condition's freshness limb and the temporal-order
limb fall to the registered study and to future work; the behavioral
condition waits on the record class whose absence is the survey's
finding). A specification paper earns its keep by stating exactly this
honestly, and the falsifier architecture of Section 1.3 attaches to what
is claimed, not to what is deferred.

\subsection{8. Threats to validity}\label{threats-to-validity}

\textbf{Grader dependence.} Documentation grading is judgment. The
mitigations are structural: a fixed protocol with the label definitions
printed, every affirmative grade carrying a verbatim quote and pinned
page, every absence bounded by named pages, and the entire grading
repeated blind with raw agreement reported (17 of 20, 55 of 60, 78 of
108) and every divergence resolution recorded. The agreement figures are
raw, over a skewed label space, and no chance-corrected statistic is
claimed; the sealed tables are exactly as auditable as their pins.

\textbf{Documentation is not behavior.} The survey grades what vendors
state, and records are known to diverge from practice
{[}\hyperref[ref-aranda2009]{8},
\hyperref[ref-hollnagel2015_whitepaper]{22}{]}. The claim is scoped
accordingly: it is a claim about what an operator can rely on the
platform to have recorded, which is what an after-the-fact justification
has to read. Where documentation overstates emission the survey errs
generous, which only strengthens the negative cells, and insulates only
them. The same generosity inflates affirmative cells, so every
affirmative count in Section 6, including the load-bearing 16 of 27
nominal-versioning defaults, is an upper bound for exactly the reason
the absences are robust.

\textbf{The measured population and the titled population, stated
precisely.} The title names agentic software engineering, and the
paper's claim to it is specific: what is measured is the
\emph{accountability structures} that setting runs on: the record
classes its platforms emit by default (the 27-platform agent class is
agentic engineering's delivery surface, surveyed exhaustively at the
documentation layer) and the realized assurance of release pipelines on
the CI/CD substrate agentic projects ship through, a frame that itself
contains agentic software (agent frameworks and agentic tools appear in
the stars-ranked stratum, so their delivery pipelines are measured at
the artifact layer). What is \emph{not} measured is any pipeline in
which the behavioral tuple is the deployable and an agent operates the
loop: the agent-class evidence is documentation, not exhaust, and no
such pipeline's exhaust is graded outside the motivating case. This is
not an accident of frame selection but the paper's own finding folded
back on its method: the agent class emits no public exhaust surface on
which a depth study could run (Section 6.1's tuple row is empty; Section
6.2's records are nominal), so realized depth is measurable today only
where records exist, which is the artifact layer of conventional
pipelines. The joint claim is therefore split-level by construction,
documentary for the agent class and measured for the artifact layer, and
the behavioral-rung gap the title's setting most needs is unmeasured
anywhere; the moment an agent platform emits an exhaust surface, the
instrument applies unchanged, and until then the split is the scope
limit of every sentence above.

\textbf{Snapshot half-life.} Platforms moved during the survey window
(one vendor rebranding mid-survey, one retirement scheduled inside the
expiry window). Every judgment is dated and pinned, the re-run
obligation is stated with its date, and the finding is framed to expire
rather than silently stale.

\textbf{The public study's visibility limits.} The exhaust visible to an
unauthenticated observer is a floor, not the whole: admin-only surfaces
(branch protection) were unreadable on all 30 repositories and are
excluded from grading by rule; empty ruleset listings are ambiguous and
graded conservatively; one collector defect (unencoded environment
names) was found, verified read-only, and recorded rather than patched
mid-study. Realized depth as measured is therefore a lower bound per
repository. The study's one post-reconciliation correction belongs in
this category and is disclosed as method: the 17 repositories first read
as ``no digests on release assets'' turned out on archive re-inspection
to publish source-only releases, a surface limit misread as a repository
property, and the affected cells were moved to undetermined by a dated
correction rather than left as measured shortfalls. The two stratum-A
shortfalls that remain are public by construction: digest-bearing assets
whose attestation lookups returned none.

\textbf{Frame and ecosystem.} One forge, one CI system, 30 repositories
in two declared strata; the frame is frozen and published, and no claim
generalizes beyond the evidence class: bounded documentation claims over
named platform populations, and a measured distribution over a declared
frame.

\textbf{Single author, with the process disclosed.} One author
adjudicated the protocol and the reconciliations; the AI-assisted
process that produced the gradings and their blind repetitions is
disclosed below, the raw pass files are archived unmodified, and every
cell is re-derivable from pinned pages and the raw archive without
trusting the author.

\subsection{9. The registered test, deliberately not reported
here}\label{the-registered-test-deliberately-not-reported-here}

The program's confirmatory arm is a prospective study at a second
operating site: predictions, analyzer, and definitions were frozen
2026-08-18 and deposited 2026-08-19 as an immutable, timestamped OSF
registration (DOI
\href{https://doi.org/10.17605/OSF.IO/D3JFV}{10.17605/OSF.IO/D3JFV}),
before any in-window promotion had run. It tests the behavior of the
four conditions under intervention, with mechanical adjudication against
pre-registered floors, and it is deliberately absent from this paper:
combining a demonstration with its test is the failure mode registration
exists to prevent. This paper's own claims are tested by different
instruments, stated where the claims are: the survey re-runs against the
pinned baseline, replication of the depth study on other frames from the
published archive and analyzer, and exhibition against the bounded
absences.

\subsection{10. Conclusion}\label{conclusion}

The traceability standards promise more than the field's instruments
check, and this paper took the promise seriously as a measurement
problem. It specified the conditions under which a promotion is
justifiable and measured, from public material only, both sides of the
field's current position: what the default record can express (a
two-class, 47-platform survey, double-graded and pinned: approval
records widely default, provenance emission default almost nowhere, and
no content-addressed behavioral-tuple identity found default on any of
the 47 platforms, in a survey of 188 double-graded cells, with nominal
versioning arriving in its place), and what declared assurance published
evidence realizes (a 30-repository distribution whose sharpest result is
that most attestation adopters declare a binding on a release surface
carrying no artifacts to check, while most adopters whose surface can be
checked realize their declaration, and both measured adopter shortfalls
sit at the binding rung). The conclusion is one sentence: the records
are structurally short of the level that can refuse a transition, the
shortfall is measurable without asking anyone, and it is dated, pinned,
and set to expire the moment the platforms make it false. That is what
it means to move from traceability to justifiability: not more records,
but records that can carry a check that refuses.

\subsection{Data and artifact
availability}\label{data-and-artifact-availability}

The research repository carries the survey record with its protocols,
raw pass files, reconciliation records and append-only pin log; the
public depth study's frozen frame, collector, hashed raw archive,
grading passes, reconciliation and deterministic analyzer; the frozen
claim sets and definitions; and the scripts that regenerate the figure
and the bibliography. The repository is private at the time of writing;
\textbf{its public release is a condition of this paper's publication,
simultaneous with the preprint.} The Site B registration is at OSF (DOI
10.17605/OSF.IO/D3JFV). What is withheld: the operating workspace
containing client-identifying material for the motivating case remains
private under the program's confidentiality rule; nothing quantitative
in this paper depends on it.

\subsection{AI usage disclosure}\label{ai-usage-disclosure}

Large language models (Anthropic Claude, including Claude Code) were
used throughout this program as engineering and drafting instruments:
for the instrument code, the analysis scripts, the adversarial-audit
tooling, the survey gradings and their blind repetitions, successive
rounds of draft review against the record, and the prose of this
manuscript, all under written protocols, with every checkable claim
verified before adoption and the dispositions archived in the program
repository. All claims, measurements, frames, and corrections were
reviewed and are the responsibility of the human author. Every citation
was verified against its primary source; the verification record is in
the program repository.

\subsection{References}\label{references}

\emph{Every entry was verified against its primary source; the per-entry
verification record is archived in the program repository.}

\begin{enumerate}
\def\labelenumi{\arabic{enumi}.}
\tightlist
\item
  \protect\phantomsection\label{ref-accountable2026}{}Accountable Agents
  in Software Engineering: An Analysis of Terms of Service and a
  Research Roadmap (2026). \emph{Proceedings of the 3rd ACM
  International Conference on AI-powered Software (AIware 2026)}.
  doi:10.1145/3805760.3814889
\item
  \protect\phantomsection\label{ref-ackerman2000}{}The Intellectual
  Challenge of CSCW: The Gap Between Social Requirements and Technical
  Feasibility (2000). \emph{Human-Computer Interaction 15(2-3):179-203}.
\item
  \protect\phantomsection\label{ref-acm2026}{}Agentic Configuration
  Management (ACM): A Reference Configuration Model for Governed Agentic
  Systems (2026). \emph{arXiv preprint}. arXiv:2608.11166
\item
  \protect\phantomsection\label{ref-adherencegap2026}{}A Large Scale
  Empirical Analysis on the Adherence Gap between Standards and Tools in
  SBOM (2026). doi:10.1145/3788692
\item
  \protect\phantomsection\label{ref-agentic2026}{}Agentic Software
  Engineering: Foundational Pillars and a Research Roadmap (2026).
  \emph{arXiv preprint (v1 2025-09-07, v3 2026-06-24)}. arXiv:2509.06216
\item
  \protect\phantomsection\label{ref-agentriskbom2026}{}AgentRiskBOM: A
  Risk-Scoping Security Bill of Materials for Agentic AI Systems (2026).
  arXiv:2606.21877
\item
  \protect\phantomsection\label{ref-agenttraces2026}{}From Agent Traces
  to Trust: A Survey of Evidence Tracing and Execution Provenance in LLM
  Agents (2026). \emph{arXiv preprint}. arXiv:2606.04990
\item
  \protect\phantomsection\label{ref-aranda2009}{}The Secret Life of
  Bugs: Going Past the Errors and Omissions in Software Repositories
  (2009). \emph{ICSE 2009 (31st Int. Conf. on Software Engineering),
  pp.~298-308}.
\item
  \protect\phantomsection\label{ref-artifactverif2026}{}Reproducibility
  is Not Enough: Artifact Verifiability in Decentralized-Build Package
  Ecosystems (2026). \emph{arXiv preprint}. arXiv:2608.18180
\item
  \protect\phantomsection\label{ref-as1105}{}AS 1105, Audit Evidence
  (2010). \emph{PCAOB Auditing Standards}.
\item
  \protect\phantomsection\label{ref-as1215}{}AS 1215, Audit
  Documentation (2004). \emph{PCAOB Auditing Standards}.
\item
  \protect\phantomsection\label{ref-auditable2026}{}Auditable Agents
  (2026). \emph{arXiv preprint; condensed version in Proceedings of the
  ACM AI Leadership Summit 2026 (Visionary Papers track)}.
  arXiv:2604.05485
\item
  \protect\phantomsection\label{ref-building2026}{}Building an Open
  AIBOM Standard in the Wild: An Experience Report on Extending SPDX 3.0
  for AI (2026). arXiv:2510.07070
\item
  \protect\phantomsection\label{ref-claimedattested2026}{}Claimed or
  Attested? A Commit-Signature Dataset and Identity Trust Tiers across
  the World of Code (2026). \emph{arXiv preprint}. arXiv:2607.06194
\item
  \protect\phantomsection\label{ref-constraints2013}{}Constraints of the
  PROV Data Model (W3C Recommendation) (2013).
  \url{https://www.w3.org/TR/prov-constraints/}
\item
  \protect\phantomsection\label{ref-cook2020_atl}{}Above the Line, Below
  the Line (2020). \emph{ACM Queue 17(6) / Communications of the ACM
  63(3):43-46}.
\item
  \protect\phantomsection\label{ref-from2026}{}From Prompt--Response to
  Goal-Directed Systems: The Evolution of Agentic AI Software (2026).
  arXiv:2602.10479
\item
  \protect\phantomsection\label{ref-goodenough2015ea}{}Eliminative
  Argumentation: A Basis for Arguing Confidence in System Properties
  (2015). \emph{Technical report CMU/SEI-2015-TR-005}.
\item
  \protect\phantomsection\label{ref-governed2026}{}Governed AI-Assisted
  Engineering: Graduated Human Oversight for Agentic Code Generation in
  Regulated Domains (2026). \emph{arXiv preprint (cs.HC), v2}.
  arXiv:2606.22484
\item
  \protect\phantomsection\label{ref-guac}{}GUAC: Graph for Understanding
  Artifact Composition (project documentation) (2026). \emph{OpenSSF
  project, guac.sh}. \url{https://guac.sh/}
\item
  \protect\phantomsection\label{ref-hawkins2011}{}A New Approach to
  Creating Clear Safety Arguments (2011). \emph{Advances in Systems
  Safety (19th Safety-Critical Systems Symposium)}.
\item
  \protect\phantomsection\label{ref-hollnagel2015_whitepaper}{}From
  Safety-I to Safety-II: A White Paper (2015). \emph{NHS England /
  Resilient Health Care Net}.
\item
  \protect\phantomsection\label{ref-intoto2019}{}in-toto: Providing
  farm-to-table guarantees for bits and bytes (2019).
  \url{https://www.usenix.org/conference/usenixsecurity19/presentation/torres-arias}
\item
  \protect\phantomsection\label{ref-intotospec2024}{}The in-toto
  framework specification (2024).
  \url{https://github.com/in-toto/docs/blob/master/in-toto-spec.md}
\item
  \protect\phantomsection\label{ref-isa230}{}ISA 230, Audit
  Documentation (2009). \emph{International Standard on Auditing
  (IAASB)}.
\item
  \protect\phantomsection\label{ref-isa500}{}ISA 500, Audit Evidence
  (2009). \emph{International Standard on Auditing (IAASB)}.
\item
  \protect\phantomsection\label{ref-iso12207}{}ISO/IEC/IEEE 12207:2026,
  Systems and software engineering -- Software life cycle processes
  (traceability, clause 3.1.69) (2026). \emph{ISO/IEC/IEEE standard;
  definition indexed as SEVOCAB headword `traceability' sense (2)}.
  \url{https://www.computer.org/sevocab}
\item
  \protect\phantomsection\label{ref-isotr18018}{}ISO/IEC TR 18018:2010,
  Information technology -- Systems and software engineering -- Guide
  for configuration management tool capabilities (traceability, clause
  3.14) (2010). \emph{ISO/IEC technical report; definition indexed as
  SEVOCAB headword `traceability' sense (3)}.
  \url{https://www.computer.org/sevocab}
\item
  \protect\phantomsection\label{ref-justified2025}{}Enabling Ethical AI:
  A case study in using Ontological Context for Justified Agentic AI
  Decisions (2025). \emph{arXiv preprint}. arXiv:2512.04822
\item
  \protect\phantomsection\label{ref-macaron2023}{}Macaron: A Logic-based
  Framework for Software Supply Chain Security Assurance (2023).
  doi:10.1145/3605770.3625213
\item
  \protect\phantomsection\label{ref-misalignment2026}{}When Model
  Release Meets Model Reuse: Producer-Consumer Misalignment in Hugging
  Face (2026). \emph{arXiv preprint}. arXiv:2607.21738
\item
  \protect\phantomsection\label{ref-mutating2026}{}Mutating the
  ``Immutable'': A Large-Scale Study of Git Tag Alterations (2026).
  arXiv:2606.31354
\item
  \protect\phantomsection\label{ref-naur1985}{}Programming as Theory
  Building (1985). \emph{Microprocessing and Microprogramming
  15(5):253-261 / repr. Computing: A Human Activity}.
\item
  \protect\phantomsection\label{ref-nistsp80053r5}{}NIST SP 800-53
  Rev.~5, Security and Privacy Controls for Information Systems and
  Organizations (control enhancement SA-8(22), Accountability and
  Traceability) (2020). \emph{NIST Special Publication 800-53 Revision
  5}. doi:10.6028/NIST.SP.800-53r5
\item
  \protect\phantomsection\label{ref-on2026}{}On Good Authority:
  Release-Authority Measurement for Registry-Mediated Package Ecosystems
  (2026). arXiv:2606.22593
\item
  \protect\phantomsection\label{ref-openlineage2026}{}OpenLineage Spec
  --- Core Lineage Model (2026).
  \url{https://github.com/OpenLineage/OpenLineage/blob/main/spec/OpenLineage.md}
\item
  \protect\phantomsection\label{ref-operationalizing2026}{}Operationalizing
  Research Software for Supply Chain Security (2026). arXiv:2601.20980
\item
  \protect\phantomsection\label{ref-prooforstop2026}{}Proof-or-Stop:
  Don't Trust the Agent, Trust the Evidence --- Loop Engineering for
  Verifiable Evidence-Gated Lifecycle Control (2026). arXiv:2607.14890
\item
  \protect\phantomsection\label{ref-provdm2013}{}PROV-DM: The PROV Data
  Model (W3C Recommendation) (2013).
  \url{https://www.w3.org/TR/prov-dm/}
\item
  \protect\phantomsection\label{ref-ptmnaming2023}{}``I see models being
  a whole other thing'': An Empirical Study of Pre-Trained Model Naming
  Conventions and A Tool for Enhancing Naming Consistency (2023).
  \emph{arXiv preprint}. arXiv:2310.01642
\item
  \protect\phantomsection\label{ref-rushby2013}{}Logic and Epistemology
  in Safety Cases (2013). \emph{SAFECOMP 2013, LNCS 8153, pp.~1-7}.
\item
  \protect\phantomsection\label{ref-rushby2015nasa}{}Understanding and
  Evaluating Assurance Cases (2015). \emph{NASA Contractor Report
  (NTRS)}.
\item
  \protect\phantomsection\label{ref-sbom_noedges2026}{}No Edges, No
  Verdict: A Large-Scale Empirical Study of Declared Dependency Graphs
  (2026). \emph{arXiv:2607.22140}.
\item
  \protect\phantomsection\label{ref-sculley2015}{}Hidden Technical Debt
  in Machine Learning Systems (2015).
  \url{https://proceedings.neurips.cc/paper_files/paper/2015/hash/86df7dcfd896fcaf2674f757a2463eba-Abstract.html}
\item
  \protect\phantomsection\label{ref-semiexecutable2026}{}The
  Semi-Executable Stack: Agentic Software Engineering and the Expanding
  Scope of SE (2026). \emph{arXiv preprint; write-up of keynote at
  Agentic Engineering 2026 workshop, Rio de Janeiro}. arXiv:2604.15468
\item
  \protect\phantomsection\label{ref-semverhf2024}{}Towards Semantic
  Versioning of Open Pre-trained Language Model Releases on Hugging Face
  (2024). \emph{arXiv preprint}. arXiv:2409.10472
\item
  \protect\phantomsection\label{ref-sigstore2022}{}Sigstore: Software
  Signing for Everybody (2022). \emph{Proceedings of the 2022 ACM SIGSAC
  Conference on Computer and Communications Security}.
  doi:10.1145/3548606.3560596
\item
  \protect\phantomsection\label{ref-silentupdates2026}{}Silent Updates:
  Measuring and Closing the Post-Deployment Disclosure Gap (2026).
  \emph{arXiv preprint}. arXiv:2608.11803
\item
  \protect\phantomsection\label{ref-slsa2023}{}SLSA v1.0 - Build
  isolation (Build L3) (2023). \emph{OpenSSF / Linux Foundation}.
  \url{https://slsa.dev/spec/v1.0/requirements}
\item
  \protect\phantomsection\label{ref-slsa2025}{}SLSA specification v1.2
  --- Terminology (2025). \url{https://slsa.dev/spec/v1.2/terminology}
\item
  \protect\phantomsection\label{ref-slsa_threats}{}SLSA Specification
  v1.0 --- Threats \& mitigations (2023). \emph{slsa.dev specification
  (v1.0)}.
\item
  \protect\phantomsection\label{ref-star_bowker2007}{}Enacting silence:
  Residual categories as a challenge for ethics, information systems,
  and communication (2007). \emph{Ethics and Information Technology
  9(4):273-280}.
\item
  \protect\phantomsection\label{ref-the2026}{}The Responsibility Vacuum:
  Organizational Failure in Scaled Agent Systems (2026). \emph{arXiv
  preprint (cs.AI), 21 Jan 2026}. arXiv:2601.15059
\item
  \protect\phantomsection\label{ref-where2026}{}Where Accountability
  Lives: Mapping Human Responsibility to Workflow Artifacts in Agentic
  Software Development (2026). \emph{arXiv preprint (cs.SE), 16 Aug
  2026; source collection at Zenodo doi:10.5281/zenodo.21965182}.
  arXiv:2608.15678
\item
  \protect\phantomsection\label{ref-woods_allspaw2020_revealing}{}Revealing
  the Critical Role of Human Performance in Software (2020). \emph{ACM
  Queue 17(6) / Communications of the ACM}.
\end{enumerate}

\end{document}